\documentclass[preprint]{aastex631}

\accepted{May 15, 2023}

\begin{document}

\title{A New LBV Candidate in M33}

\correspondingauthor{John C. Martin}
\email{jmart5@uis.edu}

\author[0000-0002-0245-508X]{John C. Martin}
\affiliation{University of Illinois Springfield}

\author[0000-0003-1720-9807]{Roberta M. Humphreys}
\affiliation{University of Minnesota}

\author[0000-0002-4582-4164]{Kerstin Weis}
\affiliation{Astronomical Institute, Faculty for Physics and Astronomy, Ruhr University Bochum}

\author[0000-0001-5126-5365]{Dominik J. Bomans}
\affiliation{Astronomical Institute, Faculty for Physics and Astronomy, Ruhr University Bochum}



\begin{abstract}

The evolutionary relationships and mechanisms governing the behavior of the wide variety of  luminous stars populating the upper H-R diagram are not well established.  Luminous blue variables (LBVs) are particularly rare, with only a few dozen identified in the Milky Way and nearby galaxies. Since 2012, the Barber Observatory Luminous Stars Survey has monitored more than 100 luminous targets in M33, including M33C-4119 which has recently undergone photometric and spectroscopic changes consistent with an S Doradus eruption of an LBV.

\end{abstract}

\keywords{Massive stars(732),Luminous blue variable stars(944),Triangulum Galaxy(1712)}


\section{}
M33C-4119 (LGGS J013312.81+303012.6) is a new Luminous Blue Variable (LBV)  candidate discovered  in an on-going survey of M33 \citep{2017AJ....154...81M}.  LBVs are rare and difficult to identify due to the infrequency of their characteristic photometric and spectroscopic variability.  Only a few dozen have been identified in the Milky Way and nearby galaxies \citep{2016ApJ...825...64H}.  Their eruptive mechanism and connection to other classes of massive stars including B[e] supergiants, warm hypergiants, and supernova impostors are poorly understood.  

M33C-4119 is a luminous OB-supergiant \citep{2014ApJ...790...48H}  in an outer spiral arm of M33 numbered B78 in Association 127 \citep{1980ApJS...44..319H}  about 12 arc minutes southwest of the galaxy center.  The designation M33C-4119 is from \citet{2015PhDT.......408B}.  It is also identified as IFM\_B 333 by \citet{1993ApJS...89...85I} and J013312.81+303012.6 by \citet{2016AJ....152...62M}.  Before 2012 it exhibited 0.1 – 0.2 mag alpha-Cygni type variations \citep{2006MNRAS.371.1405H,2016arXiv161205560C,2015PhDT.......408B} including measurements of digitized photographic plates as far back as 1968 \citep{2017MST.......408G}.  A few measurements $\sim$0.5 mag brighter than average were recorded 2001--2002 \citep{2015PhDT.......408B,2016AJ....152...62M}.   

\begin{figure}[ht!]
\plotone{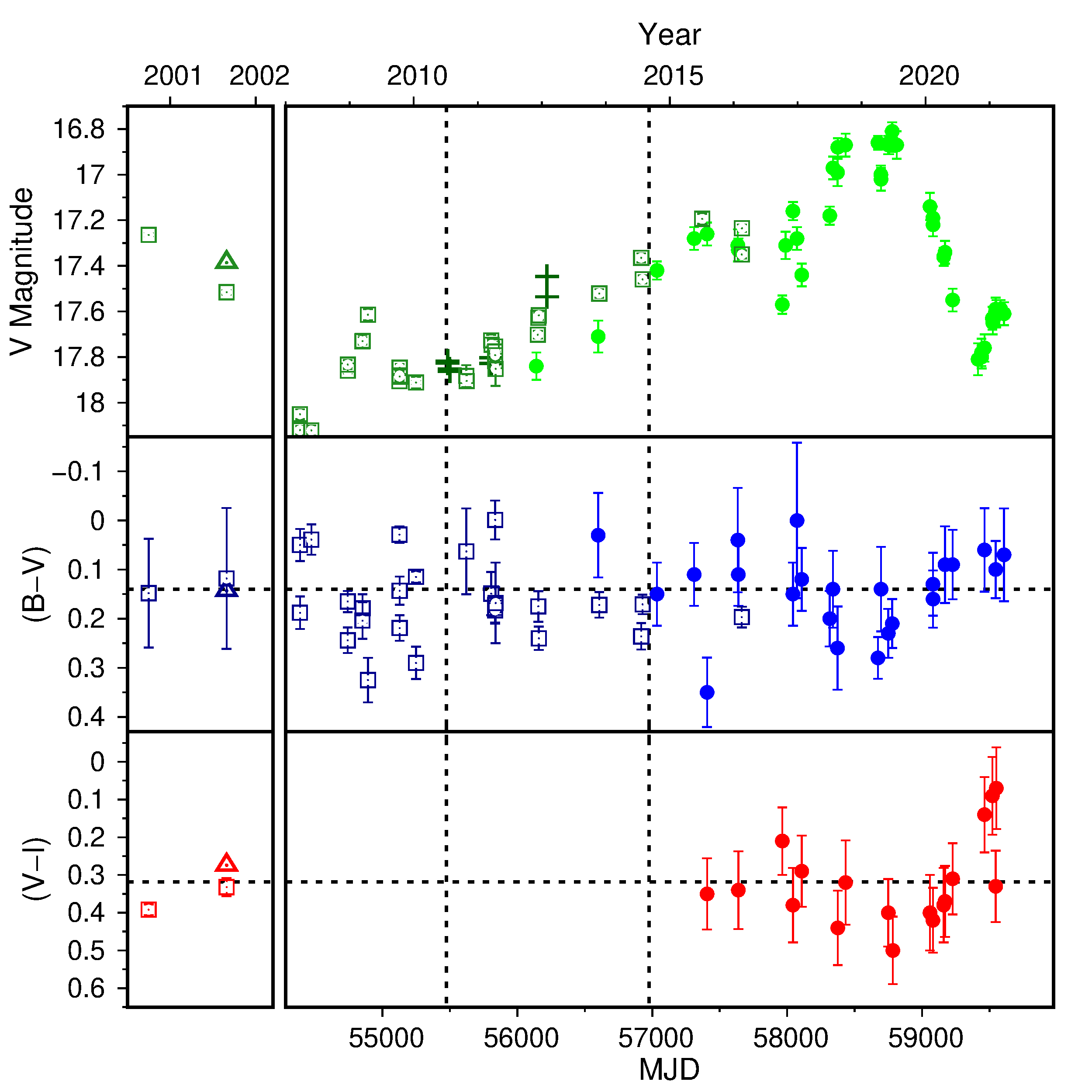}
\caption{The V magnitude and colors of M33C-4119 (LGGS J013312.81+303012.7).  The triangles are the LGGS \citep{2016AJ....152...62M}. The open squares with error bars are \citet{2015PhDT.......408B} and \citet{2017MST.......408G} CCD photometry from the Tautenburg Landessterwarte 2-m telescope.  The crosses are PanSTARRS Sloan g \citep{2016arXiv161205560C}.  The filled circles with error bars are \citet{2017AJ....154...81M} and this work.  The dashed horizontal lines in the color plots note the average value.  The dashed vertical lines mark the times of the spectra recorded in 2010 and 2014. 
\label{fig:general}}
\end{figure}

From 2012 to the present its brightness has been recorded several times a year by a BVRI CCD survey of M33 conducted with the University of Illinois Springfield Barber Observatory 20-inch telescope\citep{2017AJ....154...81M}.  From 2012 – 2018 it brightened $\approx$ 1.0 mag in all observed bands, including a rapid rise beginning in 2017.  The initial stage, 2012--2016, is confirmed by CCD photometry recorded by the Tautenburg Landessterwarte 2-m telescope \citep{2015PhDT.......408B,2017MST.......408G}. It maintained peak brightness ($\approx$ 0.5 magnitude brighter than seen previously) for more than a year before a more rapid decline to its minimum brightness by 2021 (Figure \ref{fig:general} and Table \ref{tab:phot}).  Throughout the event the star’s color was correlated with its change in brightness as expected during an S Doradus eruption, being significantly bluer/hotter when fainter and redder/cooler when visually brighter.

\begin{figure}[ht!]
\plotone{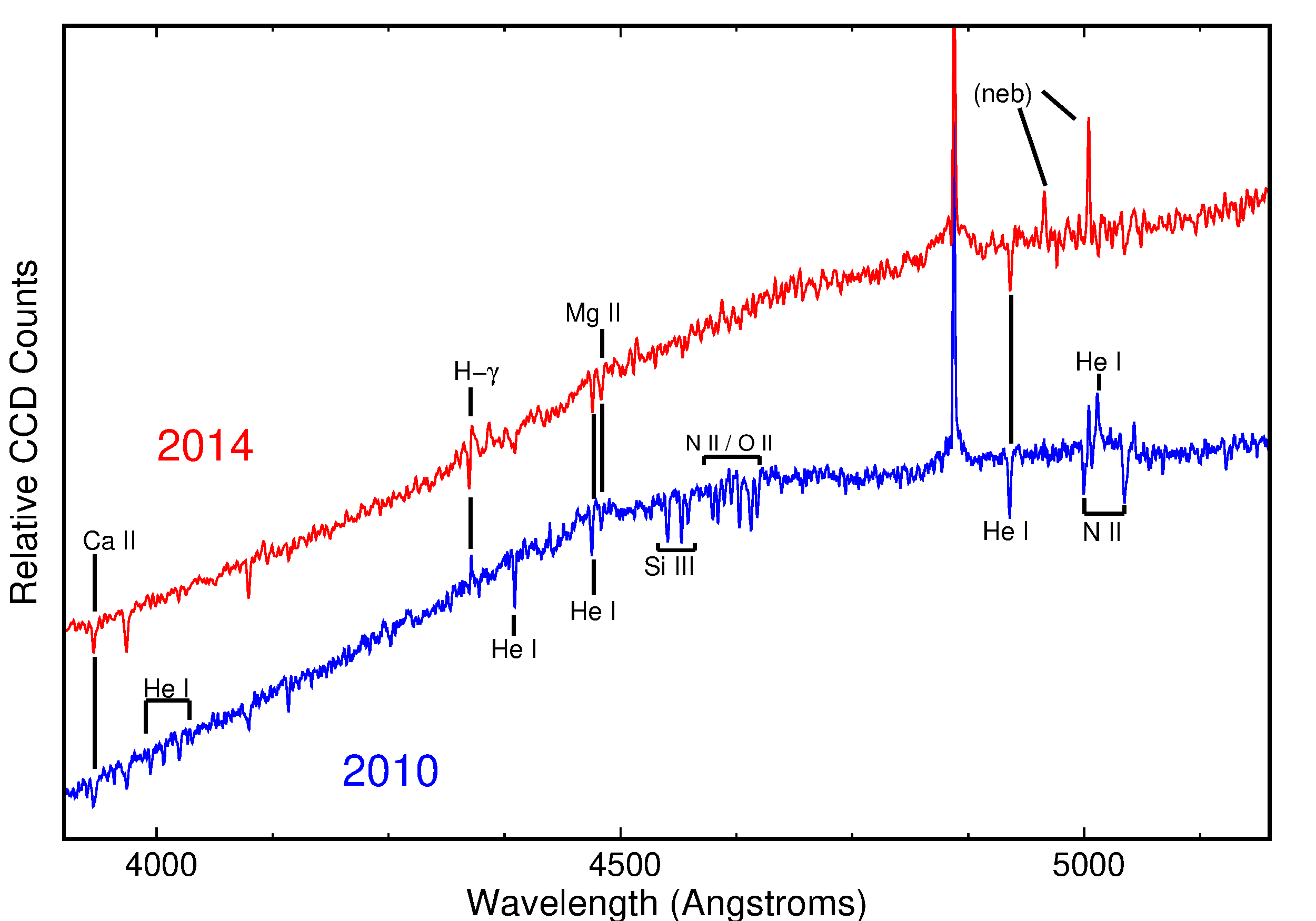}
\caption{The spectrum of M33C-4119 (LGGS J013312.81+303012.7) recorded using the blue channel of the 600-line grating of the MMT Hectospec Multi-Object Spectrograph in 2010.76 and 2014.88.
\label{fig:spectrablue}}
\end{figure}

\begin{figure}[ht!]
\plotone{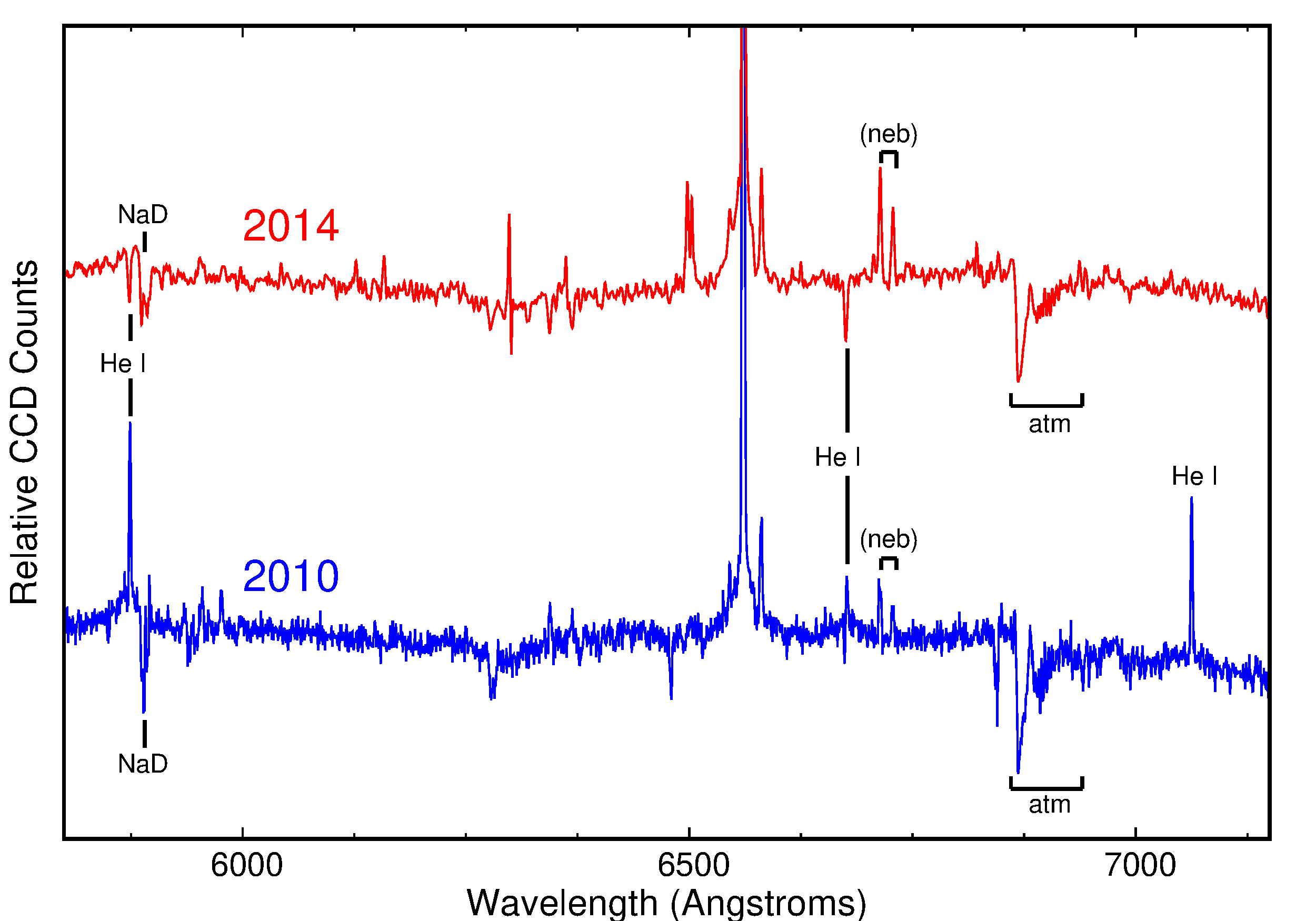}
\caption{The spectrum of M33C-4119 (LGGS J013312.81+303012.7) recorded using the red channel of the 600-line grating of the MMT Hectospec Multi-Object Spectrograph in 2010.76 and 2014.88.
\label{fig:spectrared}}
\end{figure}

Spectra of M33C-4119 were obtained with the MMT Hectospec Multi-Object Spectrograph in 2010.76 and 2014.88 with the 600-line grating in the blue and red covering 3600 to 8300A (Figure \ref{fig:spectrablue} and \ref{fig:spectrared}) and also with the LBT MODS spectrograph on 2011.75  \citep{2013ApJ...773...46H,2014ApJ...790...48H,2017ApJ...844...40H}.   The 2010 and 2011 spectra closely resemble each other. The 2014 spectrum shows a significant change and a shift to a cooler apparent temperature.  A fourth spectrum recorded in 2007.76 by \citet{2015PhDT.......408B}  using the CAFOS spectrograph on the Calar Alto 2.2-m telescope is similar in appearance to the 2010 and 2011 spectra with lower resolution and a higher noise level which affects the clear detection of weaker emission and absorption features.

Both the 2007 and 2010 spectra show a hot supergiant with a stellar wind and mass loss.  The H$\alpha$ and H$\beta$ emission lines have prominent electron scattering wings in both spectra, and in the 2010 spectrum P Cygni absorption minima are present at H$\beta$ and H$\gamma$.  In both 2007 and 2010, strong He~I emission is present at $\lambda$~5876, 6678 and 7065.  In the 2010 spectrum other He~I lines are present in absorption.  Strong absorption lines of Si~III, N~II and O~II are present along with a weak Mg~II $\lambda$~4481 line.  The absorption lines suggest an early B-type supergiant of spectral type B2 – B3 when the star was at minimum brightness.  

In 2014 when the star was halfway through its period of brightening, the He~I emission lines are replaced by absorption and the He~I lines previously present in absorption are weaker.  Absorption lines of Ca~II~K, the Na~I~D lines and Mg~II are significantly stronger, indicating a shift to cooler temperatures consistent with a late B-type spectral type ($\approx$ B8). 

\begin{figure}[ht!]
\plotone{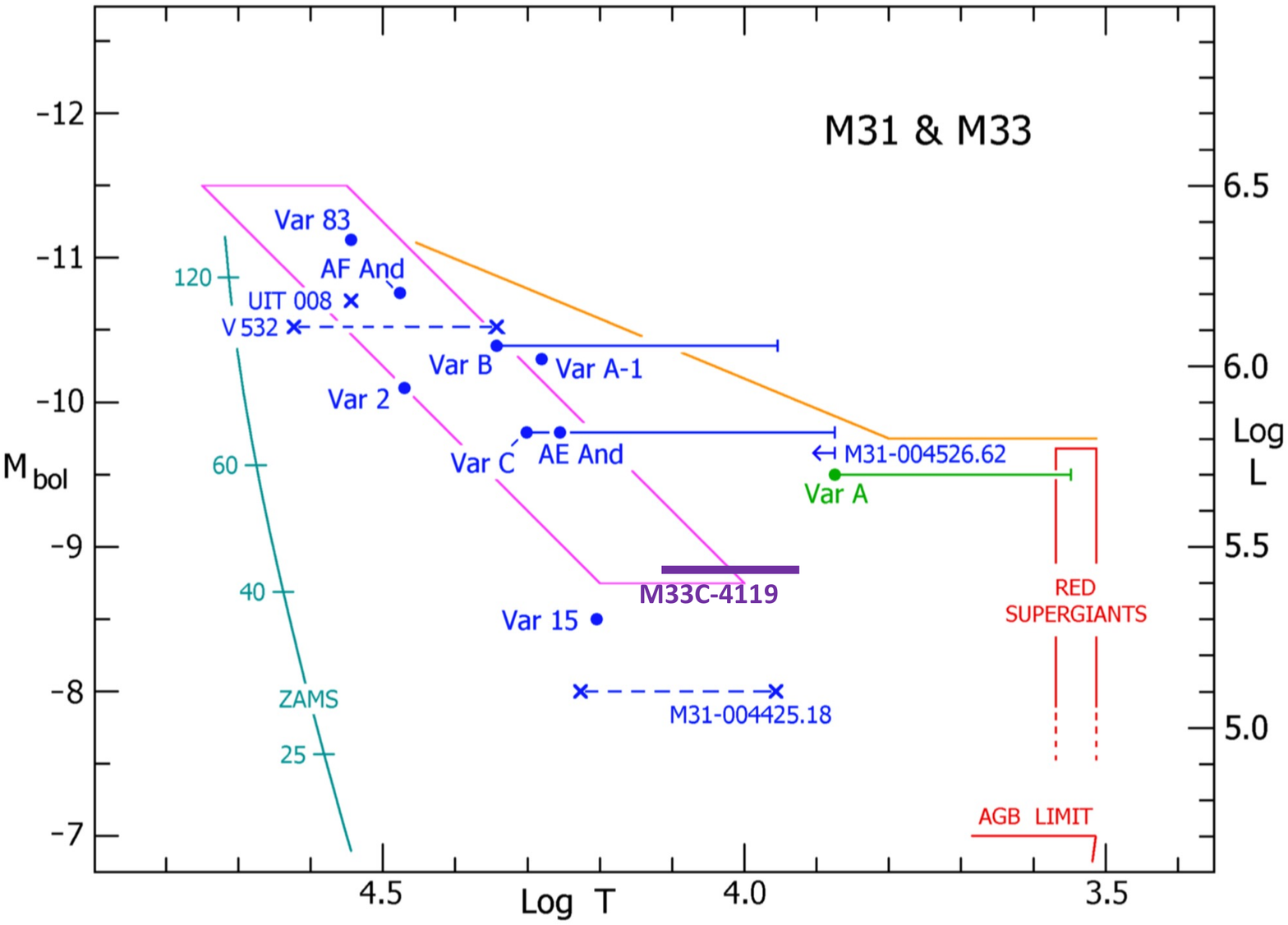}
\caption{An HRD for M31 and M33 reproduced from \citet{2016ApJ...825...64H} showing the location of M33C-4119 (LGGS J013312.81+303012.7).
\label{fig:HRD}}
\end{figure}

To estimate the star’s total bolometric luminosity and place on an HR Diagram, we determined the visual extinction. Since M33C-4119 has strong emission lines, we adopt $A_{V} = 1.10$ from two nearby OB stars.  Although we lack a spectrum at maximum light, we argue that the observed shift in color supports an equivalent  late A or F spectral type with little or no bolometric correction. A maximum V $\approx$ 16.8 mag at a distance modulus of 24.5 mag \citep{2009MNRAS.396.1287S} implies M$_{v} = M_{bol}  \approx$ -8.8 (Figure \ref{fig:HRD}.  This is also consistent with the luminosity estimated from the spectral types and brightness observed in 2010 and 2014.

LBVs are defined by S Doradus eruptive episodes characterized by a 1-2 mag increase in visual brightness accompanied by an apparent shift to cooler temperatures and  change in spectral type to  late-A to F type with little or no appreciable change in luminosity.   Many hot supergiants have emission lines and the B[e] supergiants are spectroscopically like LBVs, thus  observing an S Dor event  is the only way to confirm a star is an LBV.  The brightening of M33C-4119 is consistent with an S Dor eruption including the time scale, the rise in visual brightness, the reddening of the colors near peak brightness, and the shift to later spectral type during the brightening.  Although there is no spectrum at maximum, the case is compelling that M33C-4119 is an LBV.   

\begin{deluxetable*}{lrrr}
\tabletypesize{\scriptsize}
\tablewidth{0pt} 
\tablecaption{Recent Photometry of M33C-4119\label{tab:phot}}
\tablehead{
\colhead{MJD} & \colhead{V}& \colhead{(B-V)} & \colhead{(V-I)}\\
}
\startdata 
56140.40&$17.84\pm0.06$&&\\
56599.22&$17.71\pm0.07$&$0.03\pm0.09$&\\
57035.14&$17.42\pm0.04$&$0.15\pm0.06$&\\
57310.16&$17.28\pm0.05$&$0.11\pm0.06$&\\
57406.10&$17.26\pm0.05$&$0.35\pm0.07$&$0.35\pm0.09$\\
57634.41&$17.31\pm0.07$&$0.04\pm0.11$&\\
57638.40&$17.33\pm0.05$&$0.11\pm0.06$&$0.34\pm0.10$\\
57964.40&$17.57\pm0.04$&&$0.21\pm0.09$\\
57988.43&$17.31\pm0.06$&&\\
58043.15&$17.16\pm0.04$&$0.15\pm0.06$&$0.38\pm0.10$\\
58073.22&$17.28\pm0.05$&$0.00\pm0.16$&\\
58108.13&$17.44\pm0.05$&$0.12\pm0.06$&$0.29\pm0.09$\\
58316.35&$17.18\pm0.04$&$0.20\pm0.06$&\\
58339.40&$16.97\pm0.05$&$0.14\pm0.08$&\\
58373.15&$16.99\pm0.06$&$0.26\pm0.08$&\\
58375.16&$16.88\pm0.04$&&$0.44\pm0.10$\\
58433.02&$16.87\pm0.05$&&$0.32\pm0.11$\\
58673.38&$16.86\pm0.03$&$0.28\pm0.04$&\\
58695.39&$17.00\pm0.04$&&\\
58696.41&$17.02\pm0.05$&$0.14\pm0.09$&\\
58750.20&$16.87\pm0.04$&$0.23\pm0.05$&$0.40\pm0.09$\\
58757.17&$16.86\pm0.05$&&\\
58779.18&$16.81\pm0.04$&$0.21\pm0.05$&\\
58784.14&$16.86\pm0.04$&&$0.50\pm0.09$\\
58812.20&$16.87\pm0.06$&&\\
59059.40&$17.14\pm0.06$&&$0.40\pm0.10$\\
59081.34&$17.19\pm0.05$&$0.13\pm0.06$&$0.42\pm0.09$\\
59082.38&$17.22\pm0.05$&$0.16\pm0.06$&\\
59161.14&$17.36\pm0.04$&&$0.38\pm0.10$\\
59171.15&$17.34\pm0.05$&$0.09\pm0.08$&$0.37\pm0.09$\\
59227.14&$17.55\pm0.05$&$0.09\pm0.07$&$0.31\pm0.09$\\
59415.39&$17.81\pm0.07$&&\\
59441.42&$17.78\pm0.06$&&\\
59442.41&$17.80\pm0.05$&&\\
59463.33&$17.76\pm0.06$&$0.06\pm0.08$&$0.14\pm0.10$\\
59522.13&$17.63\pm0.05$&&$0.09\pm0.10$\\
59525.12&$17.65\pm0.05$&&\\
59546.20&$17.59\pm0.05$&$0.10\pm0.06$&$0.33\pm0.09$\\
59551.14&$17.61\pm0.06$&&$0.07\pm0.11$\\
59583.20&$17.59\pm0.04$&&\\
59608.14&$17.61\pm0.05$&$0.07\pm0.09$&\\
\enddata
\tablecomments{From survey of luminous stars in M33 described in \citet{2017AJ....154...81M}.}
\end{deluxetable*}

\begin{acknowledgments}
The UIS Barber Observatory survey of luminous stars in M33 was initiated under and supported by NSF grant AST-1108890 with additional support from the University of Illinois Springfield Henry R. Barber Astronomy Endowment funded by the people of Central Illinois.  

We also thank Brigita Burggraf and Niels Gottschling for the photometry and spectroscopy they contributed.

This work also made use of the Pan-STARRS1 Surveys (PS1) and the PS1 public science archive which has been made possible through contributions of a number of organizations credited in \citet{2016arXiv161205560C}.
\end{acknowledgments}

%





\bibliography{M33C-4119_RNAAS}{}
\bibliographystyle{aasjournal}



\end{document}